\documentclass[doublecol]{epl2}
\usepackage{amsmath,amsfonts,amssymb}
\usepackage[latin1]{inputenc}
\usepackage{graphicx}
\usepackage{color}

\title{Quasi-gaussian velocity distribution of a vibrated granular bilayer system}\shorttitle{Quasi-gaussian velocity distribution}

\author{Alexis Burdeau\and Pascal Viot}
\institute{Laboratoire de Physique
Th\'eorique de la Mati\`ere Condens\'ee, Universit\'e Pierre et Marie Curie, 4, place
Jussieu, 75252 Paris Cedex 05, France}

\abstract{
We show by using a Discrete Element Method that  a bilayer of vibrated
granular  bidisperse spheres  exhibits the striking   feature that the
horizontal velocity  distribution  of the top  layer particles   has a
quasi-Gaussian shape,  whereas that  of  the bottom  layer is far from
Gaussian.  We  examine  in   detail the  relevance of   all   physical
parameters (acceleration   of  the bottom  plate,   mass  ratio, layer
coverage).  Moreover, a microscopic analysis  of the trajectories  and
the collision statistics reveal  how   the  mechanism    of
randomization.}

\begin{document}
\maketitle

Granular  particle dynamics  and  granular  flows  are  dominated  by
dissipation  due  to  the    inelastic collisions  occurring   between
particles.  In the last twenty years,  many studies  have been carried
out on   these  non-equilibrium systems, revealing    several specific
features.  Many of the  experimental studies  have focused on vibrated
granular    media\cite{Aranson2006a}.   Among  these systems vibrated
granular  layers  are of  particular interest:  a large  variety of
patterns         occur\cite{MUS1995},           for example        two-phase
coexistence\cite{Prevost2004},     clustering\cite{Olafsen1998}     or
melting\cite{Olafsen2005}.  At  high density,  monolayer systems  display
many behaviors observed in glassformers such as a stretched intermediate
scattering        function\cite{dauchot:265701,Reis2007} and     dynamic
heterogeneities\cite{dauchot:265701}.  The way  energy  is injected into
these systems is  of   crucial importance for  their  thermostatistic
properties.

The theoretical treatment  of   energy injection in granular   systems
remains an open problem. In the framework  of kinetic theory, used for
dilute   granular flows, several   ``thermostats''  have been formally
introduced to  inject energy to the  granular particles.  They consist
of   applying external  forces to  each  particle of   the system. The
velocity distributions obtained  in these cases  are dependent on  the
way energy  is supplied.  Two types  of thermostats  have  been
extensively studied; the so-called Stochastic thermostat\cite{VE98}
where    the   force   is     a  white   noise,   and the    Gaussian
thermostat\cite{MS00} where the force is a Langevin-like force. These
two    models     both   lead     to     nearly     Gaussian  velocity
distributions. Another trick commonly used to inject
energy in kinetic theory   is to  consider  a tracer   in a  Gaussian  bath.  The  tracer
undergoes inelastic collisions with  the  bath particles, whereas  the
latter collide elastically with each  other. The velocity distribution
obtained for the tracer in  this case   is purely  Gaussian\cite{Garzo1999,MP99}. Up to now,
these  theoretical cases  have  not  been  explicitly related to
experimental situations.

Recent  experiments  clarifying   the  mechanism  of  energy
injection  in  quasi-2D     vibrated  systems have been reported. They     revealed  that
non-equilibrium steady states (NESS) can display features surprisingly
close  to those  observed   in equilibrium  systems: Prevost  {\it  et
al.}\cite{Prevost2002a} investigated  a monolayer of  steel beads
on a base plate subject  to a sinusoidal  displacement. When the plate
is rough and  the density  is low, the  velocity distribution  is very
similar  to a   Gaussian.   In   their  recent work,   Reis {\it    et
al.}\cite{Reis2007a}   have claimed to observe characteristic  features of  the
stochastic  thermostat  on the   velocity  distribution of  a vibrated
granular   monolayer.   Baxter   and  Olafsen\cite{Baxter2003}    have
experimentally  studied a  vibrated bilayer  system,  where the bottom
layer was dense and  composed of  steel beads  and  the top layer  was
composed  of  plastic beads.    The  base of  the  cell,  a horizontal
circular plate,    was   vibrated   sinusoidally  in     the  vertical
direction.  In  these experiments, the  excitation   was tuned via the
dimensionless acceleration defined as $\Gamma= A(2\pi f)^{2}/g$, where
$g$ denotes the gravity, $A$ the amplitude of the oscillations and $f$
the   frequency. The values   of $\Gamma$ used   varied from $1.75$ to
$2.25$  such that the layers were  stable. The coverage density of the
top layer, defined as the ratio of the number of particles in the layer divided by the number of particles there would be in a closed packed configuration, was varied from $c=0.2$ to $c=0.8$. The horizontal velocities of
light particles (second  layer) were monitored  to build the  velocity
distribution function. Deviations  from Gaussian are quantified by the
kurtosis $F=\frac{<v^{4}>}{<v^{2}>^{2}}$, which is equal  to $3$ for a
Gaussian distribution. The   experimental  results  showed that    the
horizontal velocity distributions of the  top layer were very close  to
Gaussian, ($|F-3|\leq 0.05)$, within  the parameters range presented above
(acceleration,   density, particle size).   Conversely, the horizontal
velocity  distribution of the   heavy-particle  layer, as well as the  vertical
velocity distributions remained strongly non Gaussian ($|F-3|\geq 1$).

The purpose  of this paper  is twofold: first to  build a  simple, but
realistic model which  captures   the observed features;  second,   to
analyze microscopic quantities to obtain insight into the role of
the  first layer  and  the  mechanisms responsible  for  this apparent
equilibrium  behavior.  In  order to examine   the robustness of  this
behavior, we  investigated  the dependence of the kinetic properties on the
mass of the  top-layer   particles.  We  also
considered the  situation where the  particles of the  first-layer are
glued to the vibrating plate, in order to  measure separately the role
of the roughness of the first layer and that of the temperature of the
first-layer particles.

Our simulation  model consists  of  a number $N_1$  of heavy spherical
particles placed at the bottom of the  simulation cell and $N_2$ light
spherical particles forming the second layer.  Periodic boundary conditions
are used in horizontal directions.   We checked  that the role of
the boundary conditions is negligible for the properties of the steady
state  by  varying the  size of  the simulation cell. In this study,the simulations were carried out on  a system with $2500$ spheres in the first layer. Initially, these  spheres are placed on a
triangular lattice, with the packing fraction of  the first
layer equal to $\eta=0.9$ (or $c=1$).

The  bottom   plate   follows  a sinusoidal   displacement   with
frequencies ranging from  $50 Hz$ to  $90 Hz$. Keeping the dimensionless acceleration constant $\Gamma=2$, namely decreasing the bottom plate vibration amplitude $A$, we do not observe variations of the kurtosis $F$ in this frequency range. In addition, we also increased the acceleration $\Gamma$ but for $\Gamma>2.5$ we observed a rapid irreversible mixing between the two layers which definitely corrupts the setup.

The collisions between spheres,  as well  as the  collisions  between spheres  and the
vibrating bottom, are inelastic.   In    addition to collisions,    the
particles are subject to a constant acceleration due to the (vertical)
gravitational   field. The  visco-elastic forces   are  modeled by the
spring-dashpot model\cite{cundallstrack79}.

Let us  consider  two spheres  labeled $1$   and $2$.  The  respective
virtual overlap $\xi$ between the two particles is given by
\begin{equation}
\xi = max(0,(R_{1}+R_{2}-|\mathbf{r_{1}}-\mathbf{r_{2}}|)).
\end{equation}
Let  $\mathbf{n}$  be the unit   vector  pointing from  the center  of
particle $1$ to the center  of particle $2$.  The simplest force along
	$\mathbf{n}$ that takes dissipation  into account is a damped
harmonic oscillator force defined as :
\begin{equation}
F_{n}=-k_{n}\xi-\gamma_{n}\dot{\xi}
\end{equation}
where $k_{n}$  is related   to  the  stiffness  of the   material, and
$\gamma_{n}$ to the dissipation.  This force model  allows one to very easily tune
 certain  quantities  in  the simulation,  especially   the
normal coefficient of restitution $e_{n}$,  which is constant for  all
collisions at all velocities in our system. Relevant values of $e_{n}$
and $t_{n}$    determine  the values    of  $k_{n}$  and $\gamma_{n}$,
independently of the velocities.  The choice of the collision duration
$t_{n}$, naturally introduces a  microscopic characteristic time.  For
the Verlet algorithm using a constant time step, we have verified that
$\Delta t =t_{n}/100 $ is a good choice for accurate results.

\begin{figure}
\centering
 \resizebox{5.5cm}{!}{\includegraphics{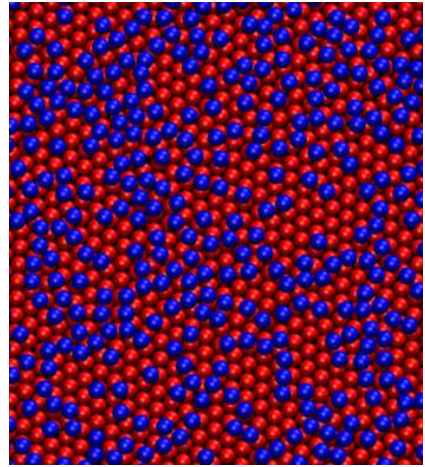}}
\caption{Snapshot of a $50\times 50$ system used in our simulations. The bottom particles (red) are densely packed on a quasi-perfect triangular lattice. The  coverage of light particles (blue) is $c=0.4$.}\label{fig:1}
 \end{figure}

In addition, we introduce a frictional force  by taking the tangential
component of the force as follows:

\begin{equation}
F_{t}= - \text{min}(|k_{t}\zeta|,|\mu F_{n}|)
\end{equation}
where $k_{t}$ is  related to the tangential  elasticity and $\zeta$ is
the    tangential  displacement   since     the   contact   was  first
established. As $k_{t}$  is related to  the stiffness of the material,
its   value  depends    on that    of  $k_{n}$    (we   used  a  ratio
$k_{t}/k_{n}=2/7$ in our simulations).

In  their  experiments, Olafsen and  Baxter\cite{Baxter2003} used
flexible dumbbells  in place of spheres  in the first layer.   For the
sake of simplicity, we have used simple spheres.  Since the density of
the first layer is high, we expect that the details of the interactions
between particles of the   first layer are   not relevant, except  for
preventing the  penetration of light  particles of the second layer into
interstitial regions, which may occur  when the system is submitted to
vibrations.

\begin{figure}
\centering
\resizebox{7.cm}{!}{\includegraphics{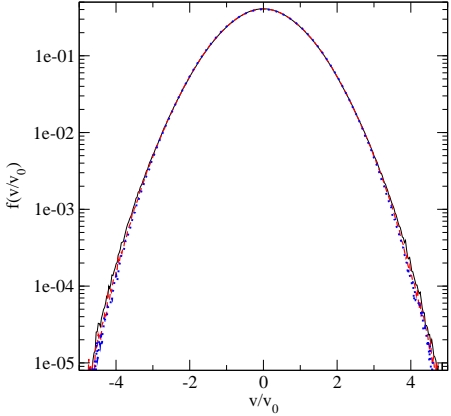}}
\caption{Renormalized velocity distributions of the light particles for different values of $e_{n} = 0.5, 0.6, 0.7$ at $\Gamma=2$ and $c=0.2$,
corresponding respectively to full (black), dashed (red),  and dotted (blue) curves.  The
magenta  dashed curve is a Gaussian   fit. The renormalizing factor is
the   root     mean  squared   velocity    $v_{0}=\sqrt{\langle  v^{2}
\rangle}$}\label{fig:2}
\end{figure}

The microscopic  parameters have been  chosen as  close as possible to
the experimental  system.  The time  step was  $\Delta t  = 10^{-5} s$,
which means that the mean duration of a collision  is $t_{n} = 10^{-3}
s$. By choosing the normal
coefficient  $e_{n_{i}}$  for the different   types of collisions, our
normal force is  then well defined.   The mass of the light  particles
is $M=22.9mg$, and the  mass of  the heavy spheres  is
taken as  one half of the mass of  the real dimers, namely $m=92.5mg$. The
diameters of light and heavy particles are the same and equal to $3mm$,
as  in the  experiment.  The  force   model  described  above  and  the
experimental setup     lead   us to  select     $5$ values of $e_{n_{i}}$
corresponding  to the different types of  collision $i$ in the system.
We  ran simulations for  various  values of  $e_{n_{i}}$ from $0.4$ to
$0.8$ (see Fig.~\ref{fig:2}). For  these values the deviations of  the
velocity  distributions from   Gaussian  are  weak, but  the  granular
temperature, defined as $T_{g}=\frac{1}{2}M
\langle v^{2}\rangle$  (where $v$ is  velocity in the   $xy$ plane)  varies by a factor 3.

The best agreement of the observed $T_{g}$ with a deviation compatible
with experiments is obtained for $e_{n}=0.5$ for  the particles of the
first layer, and   we  chose  $e_{n}=0.7$  for   the other types    of
collisions. The value $e_{n}=0.5$ for the  collisions within the first
layer  corresponds  to strongly  dissipative  collisions, but the first
layer of our system is composed of monomers instead of the dimers used
in the experiment. We assume that the dissipation occurring in a layer
of composite dimers is much more significant than in a layer of simple
spheres.   The  planar temperature  $T_{g}$  decreases with increasing
frequency  and  with increasing coverage,  as observed experimentally.
This  result  is related   to  the decrease  of  the  amplitude of the
excitation in the  first case, the  light spheres being more likely to
sit on the lattice of the first layer, and to the increase of number of
collisions in the second case.

\begin{figure}
\centering
\resizebox{7.cm}{!}{\includegraphics{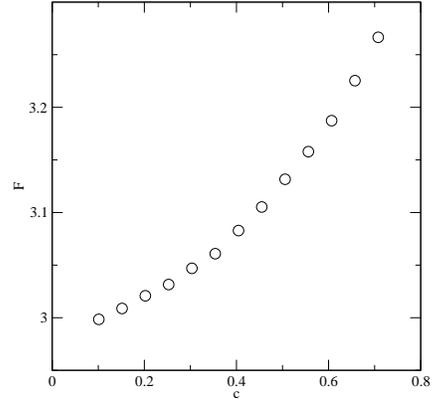}}
\caption{Kurtosis of the light particles velocity distribution as a function of the coverage of the top layer, other parameters being constant, $\Gamma=2$, $f=50Hz$. The Gaussian character remains accurate for moderate densities.}\label{fig:3}
\end{figure}

Simulations were performed by increasing the coverage of the top layer, namely increasing the number of the light particles, all others parameters being unchanged ($\Gamma=2$, $e_n$, $m$ and $M$). Figure \ref{fig:3}  shows the increase of the kurtosis with the top layer coverage. As expected, at high coverages, the deviations from  Gaussian  become significant. The remainder of our study is restricted to coverages $c\leq 0.4$.

\begin{figure}
\centering
\resizebox{7cm}{!}{\includegraphics{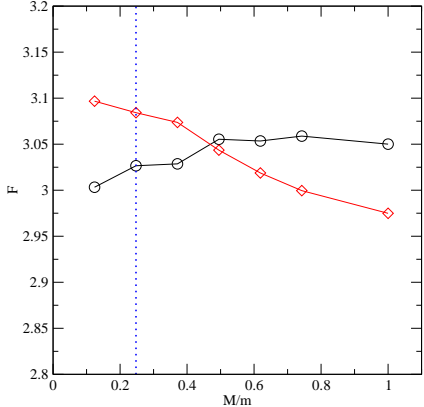}}
\caption{Kurtosis as a function of the ratio of the light particles mass to the heavy particles mass. Black circles and red diamonds correspond to $c=0.2, 0.4$ respectively. The dotted vertical line corresponds to the experimental mass ratio.}\label{fig:4}
\end{figure}

It is worth noting that with the physical parameters, ($\Gamma=2$, $c\leq0.4$),  the   first layer kurtosis of the heavy particles is always significantly higher than  that of the light particles, i.e.  $|F-3|\geq 1$,  whereas the values  of the second layer
are very close to Gaussian ones, with $|F-3|\leq 0.1$. Finally, we have also studied the influence of the mass ratio of the two species on the velocity distributions:  the kurtosis $F$ associated with the light particle velocity distribution is plotted as a function of the mass ratio $M/m$ (see Fig.\ref{fig:4}) for two values of the coverage $c=0.2,0.4$  and exhibits small variations with the mass ratio. 

\begin{figure}
\centering
\resizebox{7.cm}{!}{\includegraphics{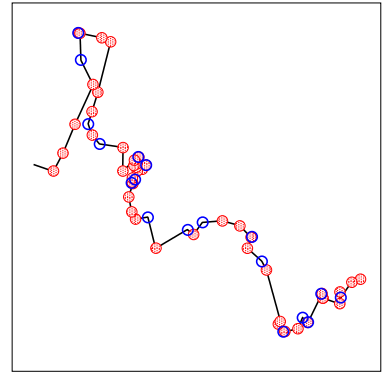}}
\caption{Trajectory of a particle of the top layer:
 collisions  with   bottom  particles  and  other top  particles are
 displayed  in  red  and blue  respectively.   Blue  collisions are on
 average separated by   several red  ones.  This randomizing   process
 explains   the nearly Gaussian character   of  the top layer velocity
 distribution. }\label{fig:5}
\end{figure}

In summary, our simulation model  captures the main  characteristics of
the velocity  distribution for different values  of the coefficient of
restitution,     as well as  for   different   densities of the  light
particles.  In  order to determine the origin  of the Gaussian profile
of the horizontal  velocity distribution, we monitored trajectories of
the light particles as well as the type of collisions occurring during
a finite duration (see Fig.~\ref{fig:5}).  Before colliding with another particle of the top layer, a light particle  undergoes one or more
collisions with the heavy  particles,  which randomizes the
horizontal   velocities, and    finally   leads to  a   quasi-Gaussian
distribution.   When   the coverage of the   top  layer increases, the
collisions   between light particles   become  more frequent than  the
collisions between  the light  and  heavy particles, and  the velocity
distribution progressively loses   its   Gaussian  character. This is illustrated in
Fig.~\ref{fig:3} which shows the variation of the kurtosis of the light particles velocity distribution with the coverage density.

When  a  collision  between  particles of  different   species occurs,
the post-collisional velocity  of the light particle
is randomized due to both the roughness of the first layer and the thermal velocity of
the  heavy particles. In order to  quantify the respective role of each
contribution,   we   performed additional   simulations  where the heavy
particles are glued to the  vibrating plate, all remaining  parameters
being unchanged: in this case, the kurtosis is about $2.9$ for a coverage density $c=0.2$. Whereas the
surface roughness of the first layer randomizes the direction of the post-collisional velocity of
the light particles,  this     sole mechanism is  not   sufficient  to lead to a very good  Gaussian  velocity profile.  When the  heavy
particles  are allowed to move, the magnitude of the light particle
velocity is modified  randomly during a collision. Combining these two mechanisms eventually allows one  to obtain a very accurate
Gaussian profile.

To  verify the homogeneity of the top layer, we have also monitored the longitudinal and transverse velocity correlation functions \cite{Prevost2004}, $C_{\parallel,\perp}(r)=\sum_{i\neq j}v_i^{\parallel,\perp}v_j^{\parallel,\perp}/N_r$ . Our results are very similar to those obtained on the experimental setup \cite{baxter:028001}, which indicate the absence of spatial correlations for the velocities of the light particles and the presence of molecular chaos.  In fact, a very high packing  in the bottom layer is a key ingredient for insuring homogeneity of the top layer:
 in a recent paper, Combs
et al\cite{combs:042301} have  shown that  if the packing  fraction of the first layer is  decreased,  holes
appear  which rapidly corrupt  the Gaussian character
of the velocity distribution. Indeed, light particles become trapped and the system loses homogeneity. In the experimental setup, the packing fraction of the first layer is high, which prevents the appearance of defects in this first layer.

We have  seen that  the system  is homogeneous, and  the non-Gaussian
character of the   velocity distribution results   from the short-time
velocity  correlations. It is possible to measure the importance of these correlations by  
considering the first-collision velocity distribution (see Ref\cite{burdeau:041305} for details):
\begin{equation}\label{eq:1}
P(z)=\langle \delta(z-{\bf v}.{\bf v}^*)\rangle
\end{equation}
where ${\bf  v}$ and  ${\bf v}^*$ are   the velocities of   a particle
before and  after a  collision.  This quantity  is different from  the
velocity correlation  function  as it  depends  on  events and not  on
time. It can be calculated analytically in a few ideal cases such as a
granular   gas  undergoing inelastic collisions, and with the assumptions of a Gaussian  velocity   distribution  and molecular chaos:  $P(z)$  has an universal feature in the sense that, for $z<0$, $P(z)$ is an  
exponential in any dimension whose analytical  expression is given as a function of physical parameters, i.e,

\begin{equation}\label{eq:2}
P(z)=P(0)e^{\frac{2}{1+e_{n}^2}\sqrt{2(1+e_{n}^2)}+1-e_{n}\frac{Mz}{T}},
\end{equation}

\begin{figure}
\centering
\resizebox{7.6cm}{!}{\includegraphics{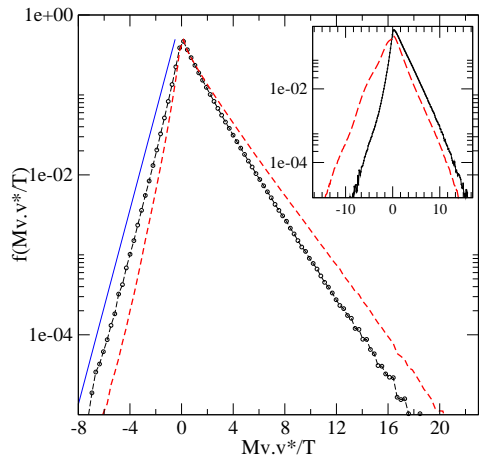}}
\caption{First-collision horizontal velocity distributions: The dotted (black) and dashed (red) curves correspond respectively to collisions between light particles and between heavy and light particles. The inset displays the first-collision three-dimensional velocity distributions. The solid (blue) curve corresponds to the exponential decay given by Eq.(\ref{eq:2}) (shifted to the left for clarity). }\label{fig:6}
\end{figure}

 This distribution is a probe for two quantities: (i) it is strongly dependent on the velocity distribution itself and
in the case   studied here  we  know  the velocity  distribution is  a
Gaussian; (ii)  it reveals  to what  extent  precollisional and post-collisional
velocities are correlated  through the collision process. By  sampling
separately the two    types   of collisions  undergone by    the  light
particles,  we     can  evaluate      and compare their
characteristics. Fig.\ref{fig:6} displays  the two distributions $P_{
lh}(z)$   and   $P_{ll}(z)$ corresponding to the horizontal velocities. The subscripts $lh$ and $ll$ denote the collisions between heavy and light particles and those between light particles, respectively.   $P_{lh}(z)$ would  be  a simple exponential law if
the light particles where reflected by  a wall but due to the randomizing effect of the bottom layer, it
is very    similar  to $P_{ll}(z)$.  This  clearly  indicates the
similarity of the collision processes  within the light beads population
and  between light  and heavy beads: in the horizontal plane,
the bottom layer behaves as a nearly Gaussian bath  in contact with the light
particles population, whereas the heavy particles velocity distribution
is  far from    being Gaussian. The theoretical result,Eq.\ref{eq:2}, is also shown for $z<0$, showing an excellent agreement with the simulation results. For completeness, we have also plotted in the inset of
Fig.\ref{fig:6} the distributions of  the three dimensional velocities. For instance,  collisions between heavy and light particles are investigated in the  decay of $P_{lh}(z)$: note that for $z<0$, $P_{lh}(z)$ is then  far from an exponential, which  is closely related to the existence of strongly non-Gaussian velocity distribution in the vertical direction.

In summary, our simple simulation model is in very good agreement with
the    experimental     results      of      Baxter     and    Olafsen
\cite{Baxter2003}. Though
simple,  the model  is  robust  and  efficient  for the simulation  of
different vibrated granular systems.  We have obtained strong evidence
that  several ingredients      are necessary  in   order to     obtain
quasi-Gaussian horizontal velocity  distribution for the  particles of
the top layer, including a rough surface that randomizes the direction
of  the  velocity of  the light  particle during a  collision with the
bottom  layer. A more    complete randomization  is  achieved if   the
amplitude  of  the velocities  of    the light particles  are  changed
``randomly'' during inter-species collisions: this occurs if the heavy
particles   are moving. The simulation   shows that these features are
robust for   a large range  of microscopic  parameters (coefficient of
restitution, mass ratio, coverage of the top layer).   We  encourage   an  extensive
experimental exploration of  this  setup, for example by  substituting
anisotropic particles for the top layer spherical particles we used.

We would like    to thank  G.W.Baxter  and J.S.Olafsen   for  fruitful
discussions and for sharing with us their experimental data as well as J. Talbot and R. Hawkins for insightful suggestions.


\end{document}